\documentclass[prl,aps,epsf]{revtex4}
\usepackage[english]{babel}
\usepackage{graphicx}
\usepackage{epsf}
\usepackage{pstricks}

\usepackage{epsfig} 
\newcommand{\wid}{.5\columnwidth}

\begin{document}

\hfill{CPHT-RR-001.0106}

\hfill{ LPT 05-81}

\hfill{hep-ph/0601176n}

\title{On exotic hybrid meson production in $\gamma^*\gamma$ collisions\\
%{\it 20 January}
}

\author{I.V.~Anikin$^{a}$,\, B.~Pire$^b$,\,
 L.~Szymanowski$^{c,d,e}$,\, O.V.~Teryaev$^{a}$,\, S.~Wallon$^e$}
\affiliation{${}^a$\,Bogoliubov Laboratory of Theoretical Physics, 141980 Dubna,
                  Russia \\
             ${}^b$\,CPHT\footnote{Unit{\'e} mixte 7644 du CNRS}, {\'E}cole
             Polytechnique, 91128 Palaiseau, France \\
             ${}^c$\,Soltan Institute for Nuclear Studies, Warsaw, Poland \\
             ${}^d$\, Universit{\'e}  de Li{\`e}ge,  B4000  Li{\`e}ge, Belgium \\
             ${}^e$\,LPT\footnote{Unit{\'e} mixte 8627 du CNRS}, Universit{\'e} Paris-Sud, 91405-Orsay, France }
\vspace{1.5cm}

\begin{abstract}

\noindent
We present a theoretical study of exotic hybrid meson ($J^{PC}=1^{-\,+}$) 
production in photon-photon collisions where one of the photons
is deeply virtual, including twist 2 and twist 3 contributions. We calculate the cross section of this process,
 which turns out
 to be large enough to imply sizeable counting rates in the present high 
 luminosity   electron-positron colliders. We
 emphasize the importance of the $\pi \eta$ decay channel for the
detection of  the hybrid meson candidate $\pi_1(1400)$ and 
calculate the cross section and the angular distribution for
$\pi\,\eta$ pair production in the unpolarized case. This angular
distribution is a useful tool for disentangling the hybrid meson
signal from the background. Finally, we 
 calculate 
 the single spin asymmetry  associated with  one initial longitudinally polarized
 lepton. 
\vspace{1pc}
\end{abstract}
\maketitle

\vspace{0.5cm}
\section{I. Introduction}
\vspace{0.5cm}

Photon-photon collisions, with one deeply virtual photon, is an excellent tool for the study of different aspects
of QCD. The main feature of such processes is that a QCD factorization theorem holds, which separates a hard partonic subprocess 
involving scattered photons from a  distribution amplitude \cite{ERBL} describing a transition of a quark--antiquark pair to a meson 
or a generalized distribution amplitude \cite{GDA} describing the  transition of a quark--antiquark pair to two- or
 three-meson states \cite{PT}. The success of this description
\cite{APT1} with respect to recent LEP data \cite{LEP} is indeed striking 
 enough to propose  such processes to be used as a tool for the discovery of new hadronic states \cite{APT2}.
 In this paper, we focus on the process where an exotic $J^{PC} = 1^{-+}$ isotriplet hybrid meson $H$
  (which may  be the $\pi_{1}(1400)$  state \cite{PDG}) 
is created in photon-photon collisions and then decays into  $\pi^0\eta$.
This exotic particle has been the subject of intensive studies for many years \cite{hybridrevue}.

In  previous papers, we have shown \cite{An04} that, contrary to naive expectations,  the longitudinally polarized
 hybrid meson possesses a leading twist distribution amplitude (DA) related to  the usual non-local quark-antiquark correlators.
We were also able to estimate the normalization of this DA, which is by no means small, with respect to the one for
usual non exotic mesons. We thus advocated that exclusive deep electroproduction processes will be able to copiously 
produce these exotic states. 
In this paper, we extend our analysis of \cite{An04} 
to $\gamma^*\,\gamma$ collisions
with production of both longitudinally and transversally polarised
hybrid meson. We calculate the hard amplitude up to the level of twist 3 and thus ignore the contributions of mass terms.
 Indeed,
the case of hybrid production and its decay products, $\pi
\eta$ pair in
electron-photon collisions is similar to the electron-proton case, with the important 
distinction that 
no unknown generalized parton distribution enters the amplitude, so that the only place where non-perturbative physics enters
is the final state DA or generalized distribution amplitudes (GDA). 
 We emphasize  the $\pi \eta$ pair production as a promising way for
detecting the hybrid meson.

Since we deal with the production of an isovector state, there is no mixing 
between quark--antiquark correlators and 
gluon--gluon correlators.  Since lepton beams are easily polarized, 
we also consider the single spin asymmetry  associated with the case 
where one of the initial lepton is longitudinally polarized while the  
polarizations of the other one are averaged over. This single spin asymmetry 
will turn out to give access to the phase difference of leading 
twist and twist 3 components of the final state GDA.

Finally let us note that in the case of the 
electron-photon collisions producing
a $\pi\eta$ pair, i.e. a state with the positive charge parity $C=+$, 
the Bremsstrahlung contribution does not exist.

\vspace{0.5cm}
\section{II. Vacuum--to--hybrid meson, vacuum--to--hadrons matrix elements and
their properties}
\vspace{0.5cm}

\subsection{The hybrid meson Distribution Amplitude}

Let us  first  consider relevant vacuum--to--hybrid meson matrix elements which enter in exclusive hard amplitudes. 
 We suppose that the hybrid meson is in the isotriplet state with
$J^{PC}=1^{-+}$ quantum numbers. A meson with such quantum numbers should be built of quarks and gluon, that is one 
should work beyond the quark-antiquark model. We have shown
\cite{An04} that 
in the case of the longitudinal hybrid meson 
polarization the leading contribution to a hard amplitude comes from
the non-local quark-antiquark correlators. 
The non-local character 
of such correlators leads to the gluonic degrees of freedom included in the gauge-invariant link.

\noindent
Let us form the light-cone basis adapted for  $\gamma^*\gamma\to H$
and $\gamma^*\gamma\to \pi \eta$ processes. We introduce the "large" and "small" vectors as 
\begin{eqnarray}
\label{lcb}
n^*=(\Lambda, {\bf 0}_T, \Lambda), \quad n=(\frac{1}{2\Lambda}, {\bf 0}_T, -\frac{1}{2\Lambda}), \quad
n^*\cdot n=1,
\end{eqnarray}
respectively, and express the photons and hybrid meson momenta via the Sudakov decomposition:
\begin{eqnarray}
\label{lcd}
q=n^*-\frac{Q^2}{2}n,\quad q^{\prime}=\frac{M^2_H+Q^2}{2}n, \quad p=n^*+\frac{M_H^2}{2}n.
\end{eqnarray}
We 
define the transverse tensor $g_{\mu\nu}^T=g_{\mu\nu}-n^*_\mu n_\nu-n_\mu n^*_\nu$.

\noindent   
Using the results of seminal studies  on the DAs of vector mesons \cite{Braun}, we write ($\bar u =1-u$): 
\begin{eqnarray}
\label{hmeV}
&&\langle H(p,\lambda) |\bar\psi(z)\gamma_\mu [z;-z]\psi(-z)|0\rangle =
\nonumber\\
&&f_H M_{H} \biggl[
p_\mu e^{(\lambda)}\cdot n
\int\limits_{0}^{1} du\, e^{i(u-\bar u) p\cdot z} \phi_1^{H}(u)
+
e_{\mu\,T}^{(\lambda)} \int\limits_{0}^{1} du\, e^{i(u-\bar u) p\cdot z} \phi_3^{H}(u)
\biggr]
\end{eqnarray}
for the vector correlator, and 
\begin{eqnarray}
\label{hmeA}
\langle H(p,\lambda) |\bar\psi(z)\gamma_\mu\gamma_5 [z;-z]\psi(-z)|0\rangle =
if_H M_{H} \varepsilon_{\mu e^{(\lambda)}_{T} p n}
\int\limits_{0}^{1} du\, e^{i(u-\bar u) p\cdot z} \phi_A^{H}(u)
\end{eqnarray}
for the axial correlator. 
 We use the following short notation : $\varepsilon_{s k m l} = 
\varepsilon_{\mu_1\mu_2\mu_3\mu_4} s_{\mu_1} k_{\mu_2} m_{\mu_3} l_{\mu_4}$. 
In eqns (\ref{hmeV}) and (\ref{hmeA}), the polarization vector $e^{(\lambda)}_\mu$ describes the spin state 
of the hybrid meson.
Due to C-charge invariance, the symmetry properties of DAs are manifested in the 
following relations:
\begin{eqnarray}
\label{symp}
\phi_1^{H}(u)=-\phi_1^{H}(1-u), \quad \phi_3^{H}(u)=-\phi_3^{H}(1-u), \quad \phi_A^{H}(u)=\phi_A^{H}(1-u).
\end{eqnarray}
Compared to the $\rho$ meson matrix elements \cite{Braun}, one can see that 
the  corresponding DAs for the exotic hybrid meson 
has different symmetry properties. 

The leading twist longitudinally polarized 
hybrid meson DA has been discussed in \cite{An04} and shown to be asymptotically equal to 
\begin{eqnarray}
\label{approxH}
\phi_{1}^{H}(u)=30 u (1-u)(1-2u)
\end{eqnarray}
and $f_{H}$ to be of the order of $50 $ MeV. In the present work, we will
not include any QCD evolution effects in GDA.

In order to be able to estimate twist 3 contributions, we have to build a model for the twist 3 DAs $\phi_3^{H}$ and 
$\phi_A^{H}$. We will not innovate on this point but restrict to the Wandzura-Wilczek parts \cite{WW}, which 
read \cite{Braun}:

 \begin{eqnarray}
    \phi_3^{WW}(u) &=& 
  \frac{1}{2}\left[
   \int_0^u dv \frac{\phi_1(v)}{v-1}
                 -  \int_u^1 dv \frac{\phi_1(v)}{ v}\right] \nonumber \\
    \phi_A^{WW}(u) &=& 
  \frac{1}{2}\left[
   \int_0^u dv \frac{\phi_1(v)}{v-1}
                 +  \int_u^1 dv \frac{\phi_1(v)}{ v}\right]\              
  \label{WW2}
  \end{eqnarray}
yielding for our choice of $\phi_{1}^H$ :

 \begin{eqnarray}
    \phi_3^{WW}(u) &=& -\,\frac {5}{2} \,(1-2u)^3
   \nonumber \\
    \phi_A^{WW}(u) &=&     \frac {5}{2} \,[1+6u(u-1)]         
  \label{WW2R}
  \end{eqnarray}

\subsection{The $\pi \eta$ Generalized Distribution Amplitude}
We now come to a discussion of the $\pi \eta$ GDA which enters the amplitude, if we try to detect the hybrid meson
through its $\pi\eta$ decay mode, which may be the dominant one for the  $\pi_{1} (1400)$ state.
Using the results of Ref. \cite{An01} we have 
\begin{eqnarray}
\label{pietameV}
&&\langle \pi^0(p_\pi)\eta(p_\eta) |
\bar\psi(-z)\gamma^{\mu}[-z;z] \psi(z)|0\rangle=
\nonumber\\
&&P^{\mu}_{\pi\eta}\int\limits_{0}^{1}du e^{i(\bar u-u)P_{\pi\eta}\cdot z}
\Phi^{(\pi\eta)}_1(u,\zeta, m_{\pi\eta}^2)+
\Delta^{\mu\,T}_{\pi\eta}\int\limits_{0}^{1}du e^{i(\bar u-u)P_{\pi\eta}\cdot z}
\Phi^{(\pi\eta)}_3(u,\zeta, m_{\pi\eta}^2)
\end{eqnarray}
for the vector correlator, and
\begin{eqnarray}
\label{pietameA}
\langle \pi^0(p_\pi)\eta(p_\eta) |
\bar\psi(-z) \gamma^{\mu}\gamma_5[-z;z] \psi(z)|0\rangle=
\varepsilon^{\mu\alpha\beta n} \Delta^{\alpha\,T}_{\pi\eta}P^{\beta}_{\pi\eta} 
\int\limits_{0}^{1}du e^{i(\bar u-u)P_{\pi\eta}\cdot z}
\Phi^{(\pi\eta)}_A(u,\zeta, m_{\pi\eta}^2)
\end{eqnarray}
for the axial correlator. In (\ref{pietameV}) and (\ref{pietameA}) the total momentum  and relative momentum of 
$\pi\eta$ pair are 
\begin{eqnarray}
P_{\pi\eta}=p_{\pi}+p_{\eta}=n^*+\frac{m^2_{\pi\eta}}{2}n, \quad 
\Delta_{\pi\eta}=p_{\pi}-p_{\eta}=(2\zeta-1 +\frac{m^2_\pi -m^2_\eta}{m^2_{\pi \eta}})P_{\pi\eta}+(1-2\zeta) m^2_{\pi\eta} n+ \Delta^T_{\pi\eta}. 
\end{eqnarray}

The $\pi \eta$ leading twist GDA has been discussed in \cite{An04} in the $J^{PC} = 1^{-+}$ channel as 
\begin{eqnarray}
\label{approxpieta}
\tilde \Phi_1^{(\pi\eta)}(u,\zeta, m_{\pi\eta}^2)=30 u (1-u)(1-2u)
\,B_{11}(m_{\pi\eta}^2) P_1(\cos\theta) ,
\end{eqnarray}
with $cos \theta = (2\zeta -1)/\beta$,
$\beta = \lambda(m^2_{\pi\eta}, m^2_{\pi}, m^2_{\eta}) / m^2_{\pi\eta}$,
 and the coefficient function $B_{11}(m_{\pi\eta}^2)$    related to the
Breit-Wigner amplitude when
$m^2_{\pi\eta}$ is in the vicinity of $ M^2_H$
as
\begin{eqnarray}
\label{B11-2}
 B_{11}(m^2_{\pi\eta})\biggl|_{m^2_{\pi\eta}\approx M^2_H}=
\frac{5}{3}\,
\frac{g_{H\pi\eta}f_H M_H \beta}{M^2_H-m^2_{\pi\eta}-i\Gamma_H \;.
  M_H}\;.
\end{eqnarray}
The formalism of GDA includes a background in a natural way,  
i.e. the contribution of final states uncorrelated with 
the main signal of hybrid meson exchange.
We will model this background as a $J=0$ contribution
(i.e. $\zeta-$independent)
 without structure in $m^2_{\pi\eta}$, and with the asymptotic 
$u-$dependence (i.e. $u(1-u)(2u-1)$).
  We know nothing of its phase but that it should be quite constant in 
the considered region. For simplicity, and contrary to the analysis
of Ref.~\cite{An04}, we do not include any $J=2$
component to the GDA. Thus our model for the leading twist
  $\pi\eta$ GDA reads

  \begin{eqnarray}
\label{approx}
\Phi_1^{(\pi\eta)}(u,\zeta, m_{\pi\eta}^2)=30 u (1-u)(1-2u)
( K e^{i\alpha} \,+\,B_{11}(m_{\pi\eta}^2) \cos\theta )\;.
\end{eqnarray}
Note that the GDAs $\Phi^{(\pi\eta)}_3(u,\zeta, m_{\pi\eta}^2), \Phi^{(\pi\eta)}_A(u,\zeta, m_{\pi\eta}^2)$ 
are  twist 3 functions; they have a part which comes from  the Wandzura-Wilczek (WW or kinematical) twist $3$ and 
another part which is usually called  genuinely twist $3$ (see, for instance \cite{An01}). The WW part reads

\begin{eqnarray}         
\label{sol_gg.2i}         
&& \Phi_3^{WW}(u,\zeta)=                   
-\frac{1}{4}\int\limits_{0}^{u}dy         
\frac{\partial_\zeta \Phi_1(y,\zeta)}{y-1}         
+\frac{1}{4}\int\limits_{u}^{1}dy         
\frac{\partial_\zeta  \Phi_1(y,\zeta)}{y} ,         
\\          
\label{sol_gg.2.2}         
&& \Phi_A^{WW}(u,\zeta)=          
-\frac{1}{4}\int\limits_{0}^{u}dy         
\frac{\partial_\zeta  \Phi_1(y,\zeta)}{y-1}         
-\frac{1}{4}\int\limits_{u}^{1}dy         
\frac{\partial_\zeta  \Phi_1(y,\zeta)}{y} \;,          
\end{eqnarray}     
which yields with our choice of the leading twist terms:
\begin{eqnarray}         
\label{sol_gg.2}         
&& \Phi_3^{WW}(u,\zeta)=  5 \,(1-2u)^3 \, \,  \frac{B_{11}(m_{\pi\eta}^2)} {2\beta}                   
\nonumber \\                  
&& \Phi_A^{WW}(u,\zeta)=   -  5 \,((1-2u)^2 \, -2u(1-u)  ) \, \, \frac{B_{11}(m_{\pi\eta}^2)} {2\beta} \;.         
\end{eqnarray}

\section{III. Amplitudes of $\gamma^*\gamma\to H$ and $\gamma^*\gamma\to \pi \eta$  processes}
\vspace{0.5cm}

\noindent
The gauge invariant expression for  the $\gamma(q^{\prime})\gamma^*(q)\to H(p)$ 
 amplitude reads 
\begin{eqnarray}         
\label{ampggH}         
T_{\mu\nu}^{\gamma\gamma^*\to H}= \frac{e^2(Q_u^2 -Q_d^2)}{\sqrt{2}}\left[         
\frac{1}{2}g_{\mu\nu}^T e^{(\lambda)}\cdot n {\bf A}_1 
+\frac{e^{(\lambda)}_{\nu\,T} \left( p+q^{\prime}\right)_{\mu}}{Q^2} {\bf A}_2\right] ,
\end{eqnarray}  
with $Q_u=2/3$ and $Q_d=-1/3$, and
where the following short notations have been introduced:
\begin{eqnarray}
\label{def1}
&&{\bf A}_1=\int\limits_{0}^{1} du E_-(u)\Phi_1(u), \quad
{\bf A}_2=\int\limits_{0}^{1} du\biggl( E_-(u) \Phi_3(u) - E_+(u) \Phi_A(u) \biggr)
\nonumber\\
&&\Phi_{1,\,3,\,A}(u)=f_H M_H \phi_{1,\,3,\,A}^H(u), \quad E_\pm(u)=\frac{1}{1-u}\pm\frac{1}{u}.
\end{eqnarray}
Such an amplitude allows calculating the production cross section and the helicity density matrix of the hybrid meson. 
The knowledge of this helicity matrix leads to a definite angular distribution for any particular decay channel.
In particular, for the process leading to the two meson $\pi \eta$ final 
state  the straightforward generalization of the amplitude derived in \cite{An01} reads
\begin{eqnarray}         
\label{ampggpieta}         
T_{\mu\nu}^{\gamma\gamma^*\to\pi\eta}=        \frac{e^2(Q_u^2 -Q_d^2)}{\sqrt{2}}\left[ 
\frac{1}{2}g_{\mu\nu}^T  {\bf A}_1^{(\pi\eta)} 
+\frac{(\Delta^{T}_{\pi\eta})_{\nu} \left( P_{\pi\eta}+q^{\prime}\right)_{\mu}}{Q^2} {\bf A}_2^{(\pi\eta)}\right] ,
\end{eqnarray}  
where, following to the notations (\ref{def1}), we introduce:
\begin{eqnarray}
\label{def2}
&&{\bf A}_1^{(\pi\eta)}=\int\limits_{0}^{1} du E_-(u)\Phi^{(\pi\eta)}_1(u)= -10\left[K\, e^{i\alpha} + B_{11}(m^2_{\pi\eta})
\cos \theta \right]\;,
\nonumber \\
&&{\bf A}_2^{(\pi\eta)}=\int\limits_{0}^{1} du\biggl( E_-(u) \Phi^{(\pi\eta)}_3(u) - E_+(u) \Phi^{(\pi\eta)}_A(u) \biggr)
= -\frac{5}{3\beta} B_{11}(m^2_{\pi\eta})
\;.
\end{eqnarray}

\noindent
Note that the amplitudes (\ref{ampggH}) and (\ref{ampggpieta}) satisfy the gauge invariance condition: 
$q_{\mu}T_{\mu\nu}=q^{\prime}_{\nu}T_{\mu\nu}=0$ provided we neglect terms  proportional to the square of the meson 
mass, which is quite natural in the light-cone formalism in the 
twist 3 approximation. 
  
  \vspace{0.5cm}
\section{IV. Cross sections}
\vspace{0.5cm}

\begin{figure}[htbp]
\begin{center}
\epsfig{file=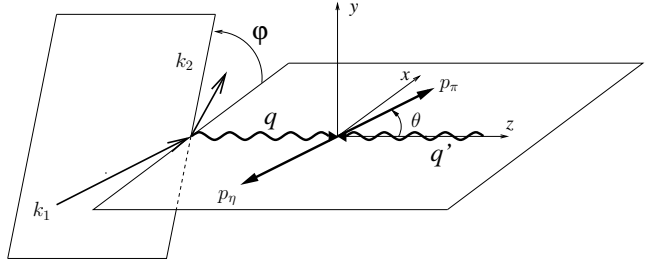,width=\wid}
\end{center}
\caption
{Kinematics of the process $e\;\gamma \to e \, \pi \, \eta\,.$
}
\label{figkinematics}
\end{figure}

The kinematics of $\gamma^*\gamma\to \pi \eta$ process is illustrated in Fig.~\ref{figkinematics}.
%%%%%%%%%%%%%%%%%%%%%%%%%%%%
%%%%%%%%%%%%%%%%%%%%%%%%%%%%
We can now calculate the  cross sections for the process $\gamma^* \gamma  \to \mbox{Hybrid meson}$. 
As for the usual treatment of a 
pseudoscalar meson \cite{BL}, it may be expressed  in terms of the transition form factor $F_{H\gamma}$, which scales like 
$1/Q^2$  up to logarithmic corrections due to the QCD evolution of the  DA (which we consistently ignore in this work). 
For an easy comparison with the well measured
cases, we define R as the ratio of 
squared amplitudes for unpolarized photons,
\begin{eqnarray}
\label{RatioR}
R = \frac {T_{\mu\nu}(\gamma \gamma^* \to H)T^{*\mu\nu}(\gamma \gamma^* \to H)}
{T_{\mu\nu}(\gamma \gamma^* \to \pi^0)T^{*\mu\nu}(\gamma \gamma^* \to \pi^0)}.
\end{eqnarray}

%%%%%%%%%%%%%%%%%%%%%%%%%%%%%
\begin{figure}[htb]
\begin{center}
\epsfig{file=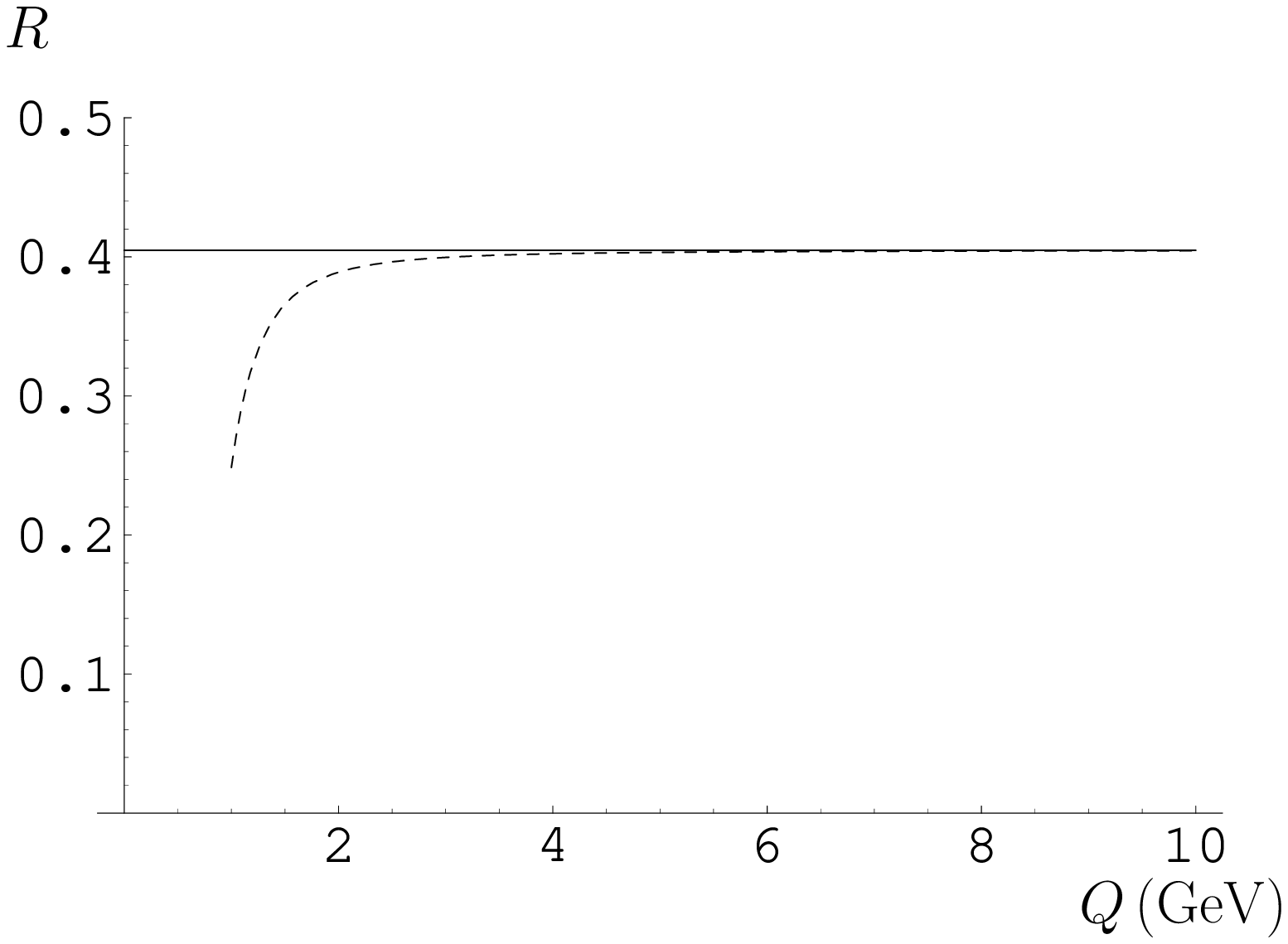,width=\wid}
\end{center}
\caption
{
The ratio $R(Q^2)$ of the squared amplitudes for $H$ and $\pi^0$ production in $\gamma^* \gamma$
collisions at leading twist and zeroth order in $\alpha_{s}$ (solid line) and including twist 3 contributions in the
numerator (dashed line).
}
\label{figR}
\end{figure}
%%%%%%%%%%%%%%%%%%%%%%%%%%%%

As shown on Fig.~\ref{figR} by the solid line, the twist 2 transition form factor 
is sizeable in the hybrid case, of the order of the corresponding 
quantity for $\pi^0$ or $\eta$ meson production. We are thus confident 
that a good detector as the ones existing in the present $e^+ e^-$ colliders will be able to detect the 
hybrid signal if one of
the decay channel has a fairly large branching ratio. To estimate the twist 3 effects, we now approximate the twist 3 part
 of the hybrid distribution amplitude in the Wandzura Wilczek way \cite{WW} as explained in Section II. 
The resulting ratio $R$ with the twist 2 and twist 3 contributions to the hybrid transition form factor  (but only the 
 twist 2 contributions to the $\pi^0$ case) is shown on Fig.~\ref{figR} by the dashed line.
The twist 3 contribution is negative and of the order of $20 \%$ when $Q \approx 1$ GeV, but quite negligible when 
$Q \geq 3$ GeV. 
Let us stress that this estimate is in the WW approximation which is by no means a proven result. 
Calculating the order of magnitude of the genuine twist 3 hybrid DA is very model dependent, and to our knowledge,
no estimate exists in the literature.

The cross section for the $e^+ e^-$ process 
\begin{eqnarray}
\label{pr2H}
e(k_1)+e(l_1)\to e(k_2)+e(l_2)+H(p).
\end{eqnarray}
for unpolarized lepton beams is easily obtained when including the leptonic parts. Note that the positive $C$ parity of the hybrid meson 
does not allow any contribution from a Bremstrahlung process. Specifying a  positive $C$ parity two body decay channel, like
$\pi^0 \eta $, we get the differential cross section for the complete
process  (see Fig.~\ref{figkinematics})
\begin{eqnarray}
\label{pr2}
e(k_1)+e(l_1)\to e(k_2)+e(l_2)+\pi(p_\pi)+\eta(p_\eta),
\end{eqnarray}
where we average (sum) over the initial (final) lepton polarizations 
and use the equivalent photon approximation (EPA) for the
 leptonic part  connected to the almost real photon, 
 \begin{equation}
  \frac{d\sigma_{ee\to ee\, \pi\eta}}{
        dQ^2\, dW^2\, d\cos\theta\, d\varphi\, dx_2} =
  \frac{\alpha}{\pi}\, \frac{1}{x_2} 
%\left(
    \frac{1+(1-x_2)^2}{2} \ln\left[
          \frac{Q'^2_{{\mathit{max}}}(x_2)}{Q'^2_{{\mathit{min}}}(x_2)}
    \right] 
%- (1-x_2) 
%\right) \,
  \frac{d\sigma_{e\gamma\to e\, \pi\eta}}{
        dQ^2\, dW^2\, d\cos\theta\, d\varphi} ,
  \label{EPA}
\end{equation}
where $W=m_{\pi\eta}$ and
 $Q'^2_{{\mathit{min}}}$ and $Q'^2_{{\mathit{max}}}$ are the
minimal and maximal virtuality of the photon $q'$, respectively. 
For a given $ee$ collider energy, the 
variables $x_2={q'p}/{l_1 p}= {s_{e\gamma}}/{s_{ee}}$ and 
$y= {q q'}/{k_1 q'}$ are 
not
independent at fixed $Q^2$ and $W^2$, since 
$y x_2 = (Q^2+W^2)/s_{ee}$. $\varphi$ is defined as being the angle between 
 leptonic and hadronic planes.
 The
lower kinematical limit $Q'^2_{{\mathit{min}}} = x_2^2\, m_e^2
/(1-x_2)$ is determined by the electron mass $m_e$, whereas
$Q'^2_{{\mathit{max}}}$ depends on experimental cuts. 
Without a precise knowledge of these cuts, we will
present results for the $e\gamma\to e\, \pi\eta$ process.
Its cross section reads
 \begin{eqnarray}
&& \frac{d\sigma_{e\gamma\to e\, \pi\eta}}{
             dQ^2\, dW^2\, d\cos\theta\, d\varphi} =  
   \frac{\alpha^3}{16\pi}\, \frac{\beta}{s_{e\gamma}^2}\,  
   \frac{1}{Q^2 }\cdot \frac{1}{2}\left(Q_u^2 -Q_d^2 \right)^2
\\
&& \Big( \frac{1+(1-y)^2}{4y^2}|A_{1}^{(\pi \eta)}|^2 +\frac{2 \bar y \beta^2 W^2}{Q^2 y^2} \sin^2\theta |A_{2}^{(\pi 
\eta)}|^2  +\frac{\sqrt{1-y} \beta W (2-y)}{Q y^2} \cos \varphi \sin\theta \, \mbox{Re} (A_1^{(\pi \eta)}A^{^{(\pi \eta)}*}_2) 
\Big) . \nonumber
 \label{gamma-gamma1}
\end{eqnarray}
To analyze the cross section, let us first define the integrated over
angles $\theta$ and $\varphi$ cross section $ d\sigma (Q^2\,, W^2)/ d Q^2 d W^2 $
%$\frac{d\sigma (Q^2\,, W^2)}{d Q^2 d W^2}$
 \begin{eqnarray}
  \frac{d\sigma (Q^2\,, W^2)}{d Q^2 d W^2}  =
 \int d\cos\theta\, d\varphi \frac{d\sigma_{e\gamma\to e\, \pi\eta}}{
             dQ^2\, dW^2\, d\cos\theta\, d\varphi} = 
\frac{25\alpha^3_{em}\beta}{72s^2_{e\gamma}Q^2}\cdot\frac{1+(1-y)^2}{y^2}\left[
  K^2 +\frac{1}{3}\,a^2(W) \,b^2(W)  \right]\;,
 \label{gamma-gamma2}
\end{eqnarray}
where we restrict ourselves to the twist 2 contribution and
\begin{equation}
\label{ab}
a(W)=\frac{5}{3}g_{H\pi\eta}f_H\beta M_H
\;,\;\;b(W)=\frac{1}{\sqrt{(W^2-M^2_H)^2+\Gamma^2_H M^2_H}}\;.
\end{equation}
In Eq.~(\ref{ab}) the coupling constant $g_{H\pi\eta}$ is related to
the branching ratio $BR=\Gamma_{H\to \pi\eta}/\Gamma_H$ of the hybrid
meson by
\begin{equation}
g^2_{H\pi\eta}= \frac{48\pi}{\beta^3M_H}\Gamma_H\,BR\;. 
\end{equation} 
Since nothing is known about the value of the $H\to \pi\eta$ BR, we
choose in our numerical analysis as a reasonable estimate
 $BR=20\%$. Let us also note that simultaneous rescaling of the
background magnitude $K$ and the BR by a coefficient $c$ amounts to a rescaling of the cross
sections by $c^2$, leaving the angular distribution $W(\theta,\varphi)$
(Eq.~(\ref{W})) unchanged. 

%%%%%%%%%%%%%%%%%%%%%%%%%%%%
\begin{figure}[h]
\begin{center}
\epsfig{file=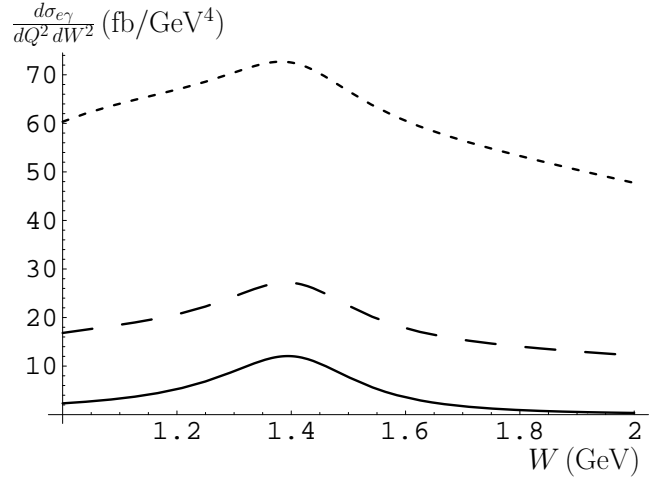,width=\wid}
\end{center}
\caption{The differential cross-section for $\pi \eta$ pair production
as a function of $W$ for $Q=3\,$GeV, $y=0.3$, for different background
magnitudes $K=0$ (solid curve), 0.5 (dashed curve) and 1
(dotted curve)
\label{figdsigKQ3}}
\end{figure}
%%%%%%%%%%%%%%%%%%%%%%%%%%%%%%%%%%%%%%%%
%%%%%%%%%%%%%%%%%%%%%%%%%%%%%%%%%%%%%%%%%%%%%%
\begin{figure}[h]
\begin{center}
\epsfig{file=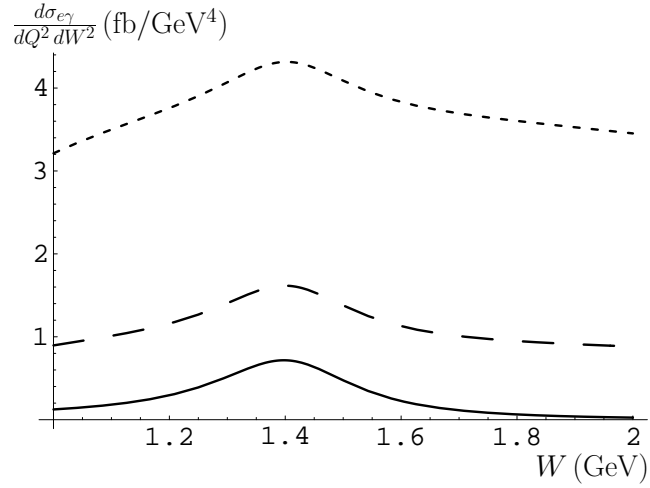,width=\wid}
\end{center}
\caption{The differential cross-section for $\pi \eta$ pair production
as a function of $W$ for $Q=5\,$GeV, $y=0.3$, for different background
magnitudes $K=0$ (solid curve), 0.5 (dashed curve) and 1
(dotted curve) 
\label{figdsigKQ5}}
\end{figure}
%%%%%%%%%%%%%%%%%%%%%%%%%%%%

This cross section does not depend on the phase $\alpha$ appearing in
our model of the $\pi\eta-$GDA, Eq.~(\ref{approx}). The plots on
Figs.~\ref{figdsigKQ3} and \ref{figdsigKQ5}
show its $W$ dependence for different magnitudes $K$ of the assumed
background, for respectively $Q=3~GeV$ and $Q=5~GeV.$ We observe that
the presence of the hybrid peak around $W=1.4$ GeV is hardly affected
when changing the magnitude $K$ of the background.
The magnitude of this signal is comparable with what has been achieved
by the L3 experiment at LEP \cite{LEP}. As expected, the comparison of
Fig.~\ref{figdsigKQ3} and Fig.~\ref{figdsigKQ5} shows
 that the magnitude of
the signal decreases when $Q$ increases. 
However, this decrease is not so dramatic due to the scaling behaviour of
the amplitude from twist two dominance.
We thus expect that the $Q$ dependence may be experimentally
studied up to a few GeV.

A more detailed test of the nature of an eventual signal may be accessed
by a study of
 the angular distribution of the $\pi \eta$ 
final state.
Using (\ref{gamma-gamma1}) and (\ref{gamma-gamma2}) supplemented by
(\ref{sol_gg.2}, \ref{ampggpieta}) and (\ref{def2}), we obtain 
\begin{equation}
\label{W}
W(\theta, \phi) =  \left( \frac{d\sigma (Q^2\,, W^2)}{d Q^2 d W^2} \right)^{-1}
\frac{d\sigma_{e\gamma\to e\, \pi\eta}}{
              dQ^2\, dW^2\, d\cos\theta\, d\varphi} =\frac{1}{4\pi}[ A + B \cos \theta 
+ C  \cos^2 \theta  +
               D \sin2\theta \cos\phi + E \sin \theta \cos\phi]\,,
\end{equation}
with the functions $A(W)$, $B(W)$, $C(W)$, $D(W,Q)$ and $E(W,Q)$:

\begin{eqnarray}
\label{ABCDE}
&&A=\frac{K^2}{K^2+\frac{1}{3}a^2 b^2} \;, \;\;\;\;\;\;
B=\frac{2Kab \,\cos \,(\gamma - \alpha) }{K^2+\frac{1}{3}a^2 b^2} 
\;, \;\;\;\;\;\;
C= \frac{a^2 b^2}{K^2+\frac{1}{3}a^2 b^2}
\nonumber \\
&&D = \frac{W}{Q} \frac{(2-y)\sqrt{1-y}}{3[1+(1-y)^2]} C
\;, \;\;\;\;\;\;
E= \frac{W}{Q} \frac{(2-y)\sqrt{1-y}}{3[1+(1-y)^2]} B \;,
\end{eqnarray}
and where $\gamma$ is the phase of the Breit-Wigner form, i.e. 
\begin{equation}
\label{gamma}
\cos \gamma= (M^2_H - W^2)\,b(W),\,\;\;\;\;\; \sin \gamma = b(W)\,\Gamma_H M_H\;.
\end{equation}

\begin{figure}[h]
\begin{center}
\epsfig{file=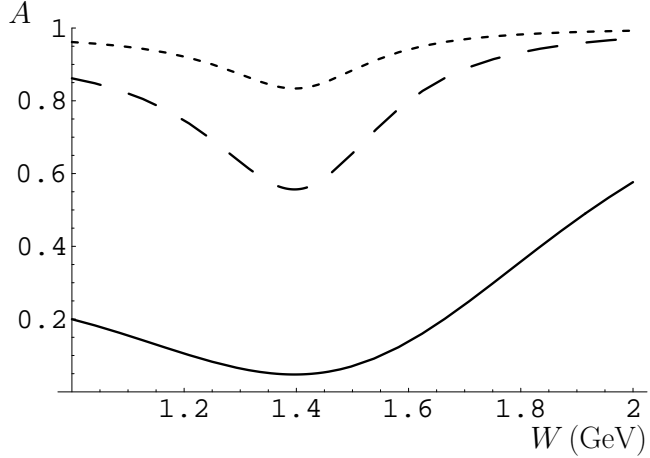,width=\wid}
\end{center}
\caption{ 
The $A$ component of the angular distribution function as a function
of the $\pi\eta$ mass  
 $W$ for $Q=3\,$GeV, $y=0.3$, for different background
magnitudes $K=0.1$ (solid curve), 0.5 (dashed curve) and 1
(dotted curve)
\label{figAK}}
\end{figure}

In (\ref{ABCDE})
 we restrict ourselves to the twist 2 contributions for the
coefficients $A(W)$, $B(W)$, $C(W)$, which  is why $A, \, B,\, C\,$ are
 $Q$ independent.
Due to the normalization of the
angular distribution $W(\theta,\varphi)$, $\int d\,\cos
\theta\,d\,\varphi W(\theta,\varphi) =1  $, the functions $A$ and $C$
are not independent but they satisfy the relation $A + C/3 =1$.
The function $A$ is displayed in Fig.~\ref{figAK}. Its shape is 
dictated by the inverse shape of the integrated cross-section shown in
Fig.~\ref{figdsigKQ3}, as seen from (\ref{ABCDE}) and (\ref{gamma-gamma2}).
This is intimately related to our main assumption
that the background is described by a $J=0$ contribution.

%%%%%%%%%%%%%%%%%%%%%%%%
%%%%%%%%%%%%%%%%%%%%%%%

\begin{figure}[h]
\begin{center}
\epsfig{file=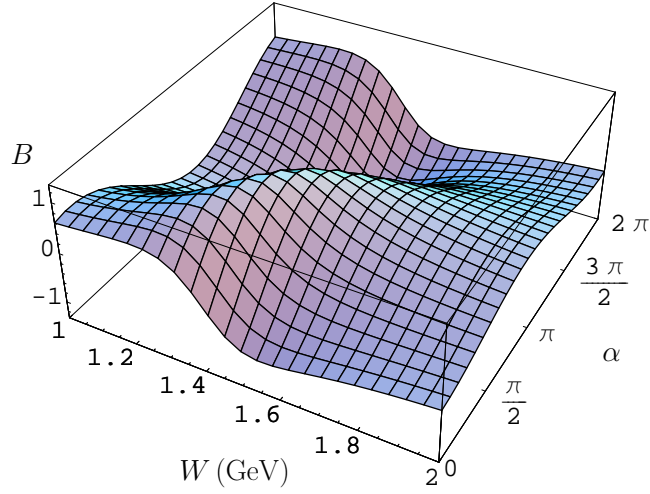,width=\wid}
\end{center}
\caption{The $B$ component of the angular distribution function as a function
of the $\pi\eta$ mass  
 $W$ and of the background phase $\alpha$ for the background magnitude
$K=1$, with $Q=3\, $GeV and $y=0.3$.
\label{figB3dK1alpha}}
\end{figure}

\begin{figure}[htb]
\begin{center}
\epsfig{file=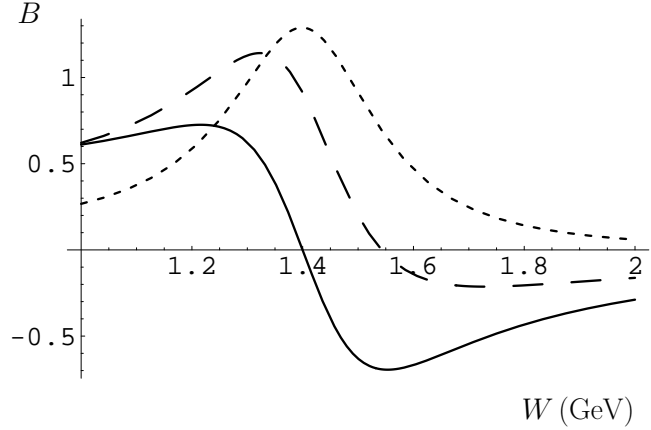,width=\wid}
\end{center}
\caption{The $B$ component of the angular distribution function as a function
of the $\pi\eta$ mass  
 $W$ for $Q=3\,$GeV, $y=0.3$ and $K=1$, for different background
phases $\alpha=0$ (solid curve), $\pi/4$ (dashed curve) and $\pi/2$
(dotted curve).
\label{figB2dK1alpha}}
\end{figure}

The $B$ coefficient is displayed in Figs.~\ref{figB3dK1alpha} and
\ref{figB2dK1alpha}, for a background magnitude $K=1$ and $y=0.3.$
It measures the interference between the background and the hybrid signal.
It is thus quite dependent on the value of the phase of the background, but its
$W$ dependence always reveals a dramatic change around the mass of the hybrid meson.
One may use it to determine more precisely this mass.

%%%%%%%%%%%%%%%%%%%%%%%%%%%%%
%%%%%%%%%%%%%%%%%%%%%%%%%%%%%%%%%%%%%%%%%
%%%%%%%%%%%%%%%%%%%%%%%%%%%%%%%%%%%%%%%%
\begin{figure}[htbp]
\begin{center}
\epsfig{file=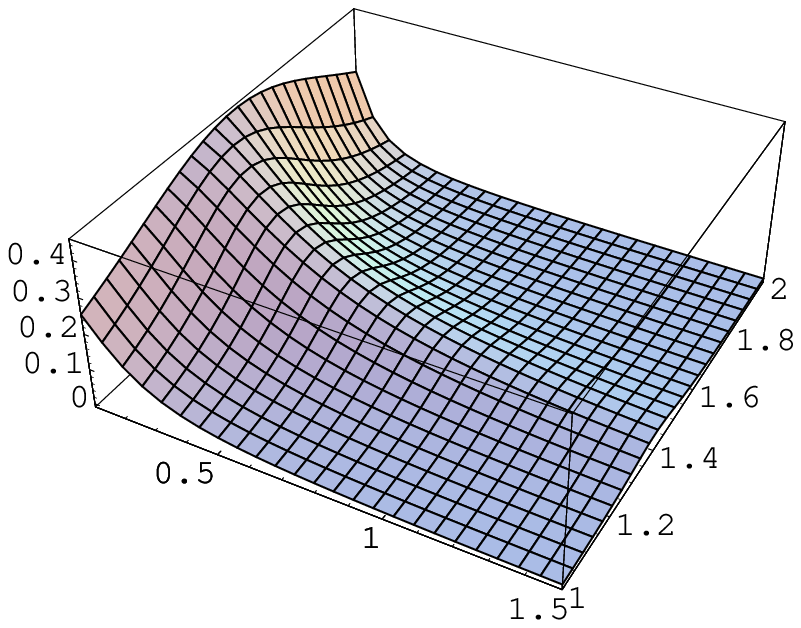,width=\wid}
\end{center}
\caption{The $D$ component of the angular distribution function as a function
of the $\pi\eta$ mass  
 $W$ and of the background magnitude $K$ for 
 $Q=3\, $GeV and $y=0.3$.
\label{figD3dKQ3}}
\end{figure}

\begin{figure}[htbp]
\begin{center}
\epsfig{file=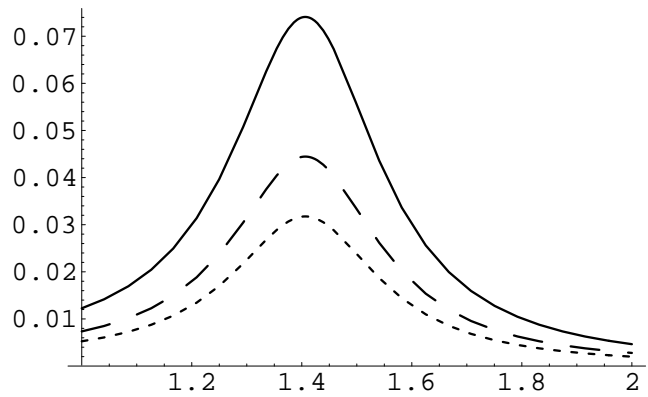,width=\wid}
\end{center}
\caption{The $D$ component of the angular distribution function as a function
of the $\pi\eta$ mass  
 $W$ for  $y=0.3$ and $K=1$, for different values of 
$Q=$3 GeV (solid curve), 5 GeV  (dashed curve) and 7 GeV  
(dotted curve).
\label{figD2dK1Q}}
\end{figure}

\begin{figure}[htbp]
\begin{center}
\epsfig{file=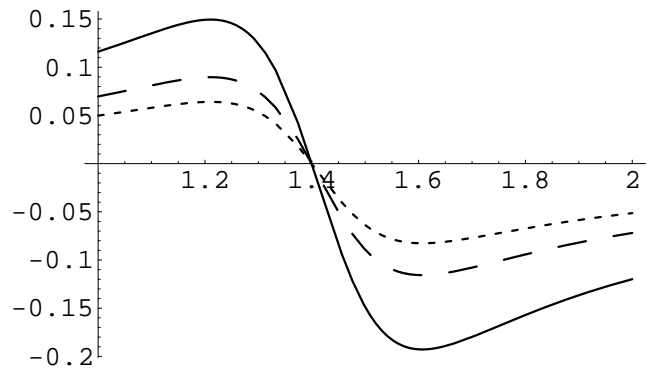,width=\wid}
\end{center}
\caption{The $E$ component of the angular distribution function as a function
of the $\pi\eta$ mass  
 $W$ for  $y=0.3$ and $K=0.5$, for different values of 
$Q=$3 GeV (solid curve), 5 GeV  (dashed curve) and 7 GeV  
(dotted curve).
\label{figEKp5Q357}}
\end{figure}

\begin{figure}[htbp]
\begin{center}
\epsfig{file=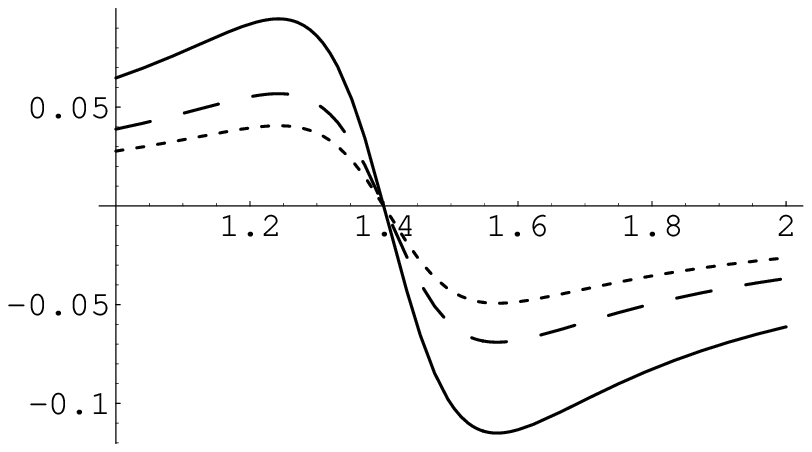,width=\wid}
\end{center}
\caption{The $E$ component of the angular distribution function as a function
of the $\pi\eta$ mass  
 $W$ for  $y=0.3$ and $K=1$, for different values of 
$Q=$3 GeV (solid curve), 5 GeV  (dashed curve) and 7 GeV  
(dotted curve).
\label{figEK1Q357}}
\end{figure}

%%%%%%%%%%%%%%%%
Since $D$ and $E$ vanish at the twist 2 level, this part of the 
 angular distribution of the final mesons is 
sensitive to the strength of the twist 3 amplitude. 
We calculate them 
in the Wandzura-Wilczek approximation
described in Section II.
The coefficient $D$ is
 independent of the phase of the background. We show on Fig.~\ref{figD3dKQ3}
its behaviour when the magnitude of the background varies, for $Q=3$ GeV and $y=0.3.$
%%%%%%%%%%%%%%%%%%%%%%%%%%
%%%%%%%%%%%%%%%%%%%%%%
On Fig.~\ref{figD2dK1Q} we show its $W$ dependence for $K=1$ and $y=.3$
and for three different values of $Q.$ Because of its proportionality with
$C,$ and thus its relation to $A,$
it is strongly peaked 
around the hybrid mass. It will only be measurable at fairly small values of
$Q.$
%%%%%%%%%%%%%%%
%%%%%%%%%%%%%%%%%
$E(W,Q)$ depends on the background phase $\alpha$, as does $B$, and on the
magnitude $K.$
We show in Figs.~\ref{figEKp5Q357} and \ref{figEK1Q357} its $W$ dependence for
$\alpha=0$
and $K=0.5$ and $K=1$ respectively.
This behaviour is similar to the one of $B$ displayed in
Fig.~\ref{figB2dK1alpha},
but its magnitude is much smaller when $Q$ is greater than 5 GeV.

%%%%%%%%%%%%%%%%%%%%%%%%%%%%%%%%%%%%%%%%%%%%%%%%%%
%%%%%%%%%%%%%%%%%%%%%%%%%%%%%%%%%%%%%%%%%%%%%%%%%%
%%%%%%%%%%%%%%%%%%%%%%%%%%%%%%%%%%%%%%%%%%%%%%%%%%%%
%%%%%%%%%%%%%%%%%%%%%%%%%%%%

\vspace{0.5cm}

\section{V. Single spin asymmetry}

\vspace{0.5cm}

\noindent

We consider now the exclusive process where a longitudinally polarized 
lepton (with helicity $h$) scatters 
on an unpolarized  photon to produce 
the lepton and the hybrid meson detected through its decay into a $\pi\eta$ pair. Such
 a process allows to define an asymmetry which is zero at the leading twist level 
but receives contributions from the interference 
 of twist 2 and twist 3 amplitudes. This asymmetry is related to the azimuthal angular  dependence of the
 polarized cross section and it is defined as   

\begin{eqnarray}
\label{ssa1}
{\cal A}_1(s_{e\gamma},Q^2,W^2; \varphi)=\frac{\int d\cos \theta_{cm} 
(d\sigma^{(\rightarrow)}-d\sigma^{(\leftarrow)})}
{\int d\cos \theta_{cm}( d\sigma^{(\rightarrow)} + d\sigma^{(\leftarrow)})}\;,
\end{eqnarray}
where we denote by $ d\sigma^{(\rightarrow)}$ the differential cross section 
${d\sigma^{(h=1)}_{e\gamma\to e\pi\eta}}/{dW^2 dQ^2 d\cos\theta_{cm} d\varphi}$.

Note that the denominator (we restrict ourselves to the dominant twist $2$ component) can be expressed through the 
unpolarized differential cross section defined in Eq.~(\ref{gamma-gamma2})
\begin{eqnarray}
\label{xsecunp}
\int\limits_0^{2\pi}d\,\varphi\,\int\limits_{-1}^1 d\cos \theta_{cms}  \;\left(   d\sigma^{(\rightarrow)} + d\sigma^{(\leftarrow)}
\right)= 2 \, \frac{d\sigma_{e \gamma} (Q^2\,, W^2)}{d Q^2 d W^2}\,.
\end{eqnarray} 

The asymmetry (\ref{ssa1}) reads:

\begin{eqnarray}
\label{ssa-f1}
&&{\cal A}_1(s_{e\gamma},Q^2,W^2; \varphi)=
\frac{
\int\limits_{0}^{\pi} d \theta_{cm} \, \sin\theta_{cm} \,{\cal N}(\theta_{cm},\,Q,\,W,\, \varphi)}
{2\int\limits_{0}^{\pi} d\theta_{cm}\sin\theta_{cm} \, {\cal D}(\theta_{cm})}
\end{eqnarray}
where
\begin{eqnarray}
\label{ssa-num1}
{\cal N}(\theta_{cm},\,Q,\,W,\, \varphi)=\frac{4}{Q^6} 
\varepsilon_{k_1 q^{\prime} k_2 \Delta^{(\pi\eta)}_T}\, 
{\rm Im}({\bf A}_1^{(\pi\eta)} {\bf A}^{*\,(\pi\eta)}_2),
\quad {\cal D}(\theta_{cm})=\frac{1}{Q^2}\frac{1+(1-y)^2}{4y^2} | {\bf A}_1^{(\pi\eta)}|^2.
\end{eqnarray}

\begin{figure}[h]
$$\includegraphics[width=10cm]{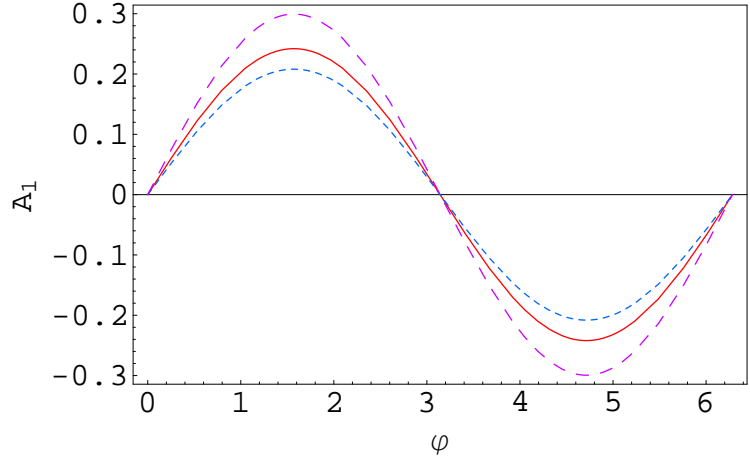}$$
\caption{
The single spin asymmetry ${\cal A}_1$ as function of $\varphi=(0,\,2\pi)$. Values of parameters:
$W=1.4\, {\rm GeV}, \, Q^2=5.0\, {\rm GeV}^2,\, s_{e\gamma}=10
  \,{\rm GeV}^2,\, \alpha=0.$
The solid line  corresponds to $K=0.8$, the short-dashed line to
$K=1.0$, the long-dashed line to $K=0.5$.}
\label{SSA1}
\end{figure}

The crucial dynamical quantity probed by this asymmetry is thus 
 ${\rm Im}({\bf A}_1^{(\pi\eta)} {\bf A}^{*\,(\pi\eta)}_2)$. It depends on the phase
 structure of the amplitudes, and may be written as
 \begin{eqnarray}
\label{ima1a2}
{\rm Im} ({\bf A}^{(\pi\eta)}_1 {\bf A}^{*\,(\pi\eta)}_2)=
\frac{50}{3}\,
 \, \frac{a(W)\;b(W)\;K}{\beta} \sin (\alpha -\gamma) \;\;,
\end{eqnarray}
where $a(W)$, $b(W)$ and $\gamma$ are defined by equations (\ref{ab}) and (\ref{gamma}).
This quantity depends much on the unknown background phase
 $\alpha$. On Fig. \ref{SSA1} , we present our result with the choice $\alpha =0$.  The resulting 
 asymmetry is sizeable and should be measurable.
%%%%%%%%%%%%%%%%%%%%%%%%%%%%%%%%%%%%%%%%%%% 
%%%%%%%%%%%%%%%%%%%%%%%%%%%%%%%%%%%%%%%%%%%%%%%%%%

\begin{figure}[h]
$$\includegraphics[width=10cm]{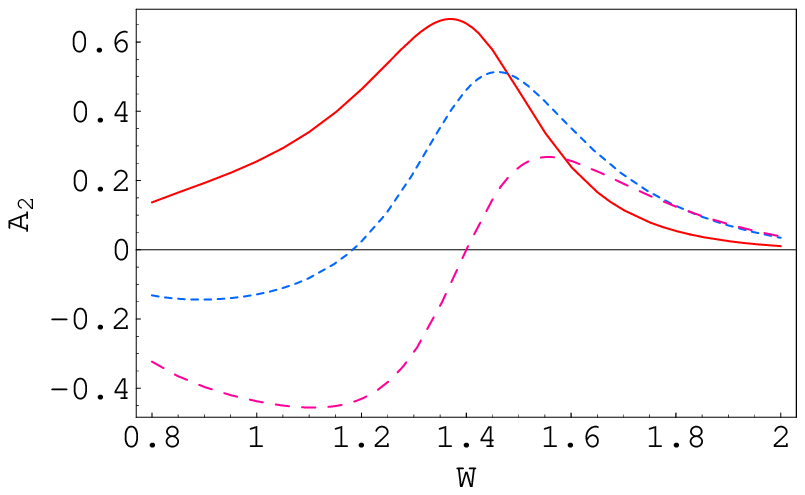}$$
\caption{
The integrated single spin asymmetry ${\cal A}_2$  as a function of the $\pi\eta$ 
invariant mass.
Values of parameters: $Q^2=5.0\, {\rm GeV}^2,s_{e\gamma}=10
  \,{\rm GeV}^2,\, K=1.0 $ , the solid line corresponds 
to $\alpha=0$, the short-dashed line to $\alpha=\pi/4$,
the long-dashed line to $\alpha=\pi/2$.}
\label{SSA2}
\end{figure}

\begin{figure}[h]
$$\includegraphics[width=10cm]{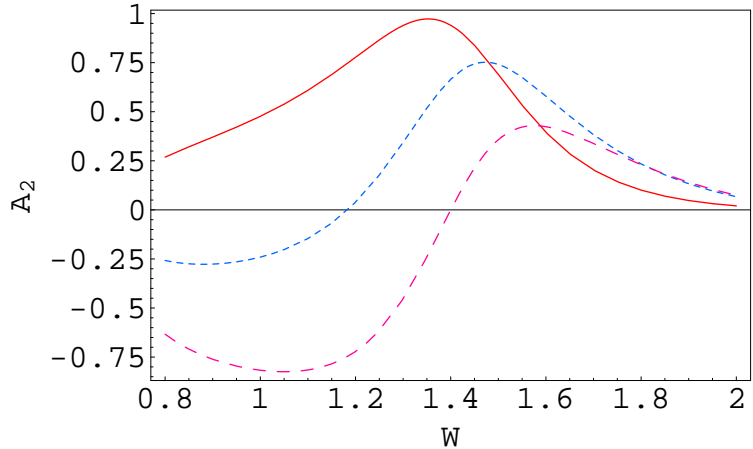}$$
\caption{
The integrated single spin asymmetry ${\cal A}_2$  as a function of the $\pi\eta$ 
invariant mass.
Values of parameters: $Q^2=5.0\, {\rm GeV}^2,s_{e\gamma}=10
  \,{\rm GeV}^2,\, K=0.5 $ , the solid line corresponds
to $\alpha=0,$ the short-dashed line to $\alpha=\pi/4$,
the long-dashed line to $\alpha=\pi/2$.}
\label{SSA22}
\end{figure}

In order to get more statistics, one may define an integrated asymmetry : 
\begin{eqnarray}
\label{ssa2}
{\cal A}_2(s_{e\gamma},Q^2,W^2)=\frac{2\pi \;\int d(\cos \theta_{cm})\, 
\int\limits_{0}^{2\pi} d\varphi\, \sin\varphi 
(d\sigma^{(\rightarrow)}-d\sigma^{(\leftarrow)})}
{\int d(\cos \theta_{cm})\int\limits_{0}^{2\pi} d\varphi\,
( d\sigma^{(\rightarrow)} + d\sigma^{(\leftarrow)})}.
\end{eqnarray}
which may be used as a probe of the $\pi \eta $ invariant mass dependence.
We show on Figs.~\ref{SSA2} and  \ref{SSA22} the resulting $W$ dependence for two choices of the background magnitude $K$. 
%%%%%%%%%%%%%%%%%%%%%%%%%%%%%%%%%%%%%%%
%%%%%%%%%%%%%%%%%%%%%%%%%%%%%%%%%%%%%%%%%%%%%
%%%%%%%%%%%%%%%%%%%%%%%%%%%%%%%%%%%%%%%%%%%%%%%%
%%%%%%%%%%%%%%%%%%%%%%%%%%%%%%%%%%%%%%%%%%%%%%%%%%%

\section{Conclusion}
This theoretical study shows that if a hybrid meson exists with $J^{PC} =
1^{-+}$ around 1.4 GeV and with a sizeable branching ratio to $\pi-\eta,$ 
much can be learned about it from the experimental observation of $\gamma^* \gamma$
reactions and the precise study of the $\pi-\eta$ final state. 
The magnitude of the cross section that we obtain in our model of the
$\pi-\eta$
Generalized Distribution Amplitude indicates that
 present detectors at current $e^+ e^-$ colliders are able to get good statistics on these reactions, provided the tagging 
 procedure is efficient.

\vspace{0.5cm}
{\bf Acknowledgements}
\noindent We acknowledge useful discussions with J.P. Lansberg, R. Pasechnik  and
M.V. Polyakov.
This work is supported in part by  RFBR (Grant 03-02-16816) and by Polish
Grant 1 P03B 028 28.
The work of B.~P., L.~Sz. and S. W. is partially supported by the French-Polish scientific 
agreement Polonium and the Joint Research Activity "Generalised Parton 
Distributions" of the european I3 program Hadronic Physics, contract RII3-CT-2004-506078. 
L.~Sz. is a Visiting Fellow of the Fonds National pour la Recherche
Scientifique (Belgium).


\vspace{0.5cm}

\begin{thebibliography}{99}




\bibitem{ERBL} G.P. Lepage and S.J. Brodsky, Phys.\ Lett.\ {\bf B87},
359 (1979); A.V. Efremov and A.V. Radyushkin, Phys.\ Lett.\ {\bf B94},
245 (1980).

\bibitem{GDA}
D.~M{\"u}ller {\it et al.},
% D.~Robaschik, B.~Geyer, F.~M.~Dittes, J.~Ho\v{r}ej\v{s}i,
Fortsch.\ Phys.\  {\bf 42},  101 (1994);
%%CITATION = HEP-PH 9812448;%%
M. Diehl, T. Gousset, B. Pire and O.V. Teryaev, Phys.\
Rev.\ Lett.\ {\bf 81}, 1782 (1998).
%%CITATION = HEP-PH 9805380;%%


\bibitem{PT}       
B.~Pire and O.~V.~Teryaev,
  %``Single spin asymmetries in gamma* gamma $\to$ pi pi pi at large Q**2,''
  Phys.\ Lett.\ B {\bf 496} (2000) 76.
  %%CITATION = HEP-PH 0007014;%%

\bibitem{APT1}
I.~V.~Anikin, B.~Pire and O.~V.~Teryaev,
  %``On gamma gamma* production of two rho0 mesons,''
  Phys.\ Rev.\ D {\bf 69} (2004) 014018.
  %%CITATION = HEP-PH 0307059;%%

\bibitem{LEP}
 P.~Achard {\it et al.}  [L3 Collaboration],
  Phys.\ Lett.\ B {\bf 568} (2003) 11,  Phys.\ Lett.\ B {\bf 597} (2004) 26, 
  Phys.\ Lett.\ B {\bf 604} (2004) 48 and  Phys.\ Lett.\ B {\bf 615} (2005) 19.
  %%CITATION = HEP-EX 0504016;%%
  %%CITATION = HEP-EX 0410073;%%
  %%CITATION = HEP-EX 0407020;%%
  %%CITATION = HEP-EX 0305082;%%

\bibitem{APT2}
I.~V.~Anikin, B.~Pire and O.~V.~Teryaev,
  %``Search for isotensor exotic meson and twist 4 contribution to gamma* gamma
  %$\to$ rho rho,''
  Phys.\ Lett.\ B {\bf 626} (2005) 86.
  %%CITATION = HEP-PH 0506277;%%
  
  \bibitem{PDG}
 S.~Eidelman {\it et al.}  [Particle Data Group],
  %``Review of particle physics,''
  Phys.\ Lett.\ B {\bf 592} (2004) 1.
  %%CITATION = PHLTA,B592,1;%%



%\cite{Chanowitz:1986gg}
\bibitem{hybridrevue}
 M.~S.~Chanowitz and S.~R.~Sharpe,
  %``Hybrids: Mixed States Of Quarks And Gluons,''
  Nucl.\ Phys.\ B {\bf 222} (1983) 211
  [Erratum-ibid.\ B {\bf 228} (1983) 588];
  R.~L.~Jaffe, K.~Johnson and Z.~Ryzak,
  %``Qualitative Features Of The Glueball Spectrum,''%\cite{Jaffe:1985qp}
  Annals Phys.\  {\bf 168} (1986) 344; 
  M.~S.~Chanowitz,
  %``An Exotic Quark - Gluon Hybrid At 1420-Mev?,''
  Phys.\ Lett.\ B {\bf 187} (1987) 409;
  %%CITATION = PHLTA,B187,409;%%
  A.~Le Yaouanc, L.~Oliver, O.~Pene, J.~C.~Raynal and S.~Ono,
  %``Q Anti-Q G Hybrid Mesons In Psi $\to$ Gamma + Hadrons,''
  Z.\ Phys.\ C {\bf 28} (1985) 309;
  %%CITATION = ZEPYA,C28,309;%%
%\cite{Close:1994pr}
  F.~E.~Close and P.~R.~Page,
  %``The Photoproduction of hybrid mesons from CEBAF to HERA,''
  Phys.\ Rev.\ D {\bf 52} (1995) 1706
  [arXiv:hep-ph/9412301];
  S.~Godfrey and J.~Napolitano,
  %``Light meson spectroscopy,''
  Rev.\ Mod.\ Phys.\  {\bf 71} (1999) 1411
  [arXiv:hep-ph/9811410];
  %%CITATION = HEP-PH 9811410;%%
S.~Godfrey,
  %``The phenomenology of glueball and hybrid mesons,''
  arXiv:hep-ph/0211464;
  %%CITATION = HEP-PH 0211464;%%
  %%CITATION = HEP-PH 9412301;%%
  F.~E.~Close and J.~J.~Dudek,
  %``Electroweak production of hybrid mesons in a flux-tube simulation of
  %lattice QCD,''
  Phys.\ Rev.\ Lett.\  {\bf 91} (2003) 142001
  [arXiv:hep-ph/0304243] and 
  %%CITATION = HEP-PH 0304243;%%
  %``Hybrid meson production by electromagnetic and weak interactions in a
  %flux-tube simulation of lattice QCD,''
  Phys.\ Rev.\ D {\bf 69} (2004) 034010
  [arXiv:hep-ph/0308098].
  %%CITATION = HEP-PH 0308098;%%






          
\bibitem{An04}
I.~V.~Anikin, B.~Pire, L.~Szymanowski, O.~V.~Teryaev and S.~Wallon,
  %``Deep electroproduction of exotic hybrid mesons,''
  Phys.\ Rev.\ D {\bf 70}, 011501 (2004)
  [arXiv:hep-ph/0401130];
  %%CITATION = HEP-PH 0401130;%%
%``Exotic hybrid mesons in hard electroproduction,''
  Phys.\ Rev.\ D {\bf 71}, 034021 (2005)
  [arXiv:hep-ph/0411407];
  %%CITATION = HEP-PH 0411407;%%
%``pi eta pair hard electroproduction and exotic hybrid mesons,''
  Nucl.\ Phys.\ A {\bf 755}, 561 (2005)
  [arXiv:hep-ph/0501119];
  %%CITATION = HEP-PH 0501119;%%
%``Hard production of exotic hybrid mesons,''
  arXiv:hep-ph/0509245;
  %%CITATION = HEP-PH 0509245;%%
%``Probing the partonic structure of exotic particles in hard
  %electroproduction,''
  AIP Conf.\ Proc.\  {\bf 775}, 51 (2005).
  %%CITATION = APCPC,775,51;%%

\bibitem{Braun}
 P.~Ball and V.~M.~Braun,
  %``The $\rho$ Meson Light-Cone Distribution Amplitudes of Leading Twist Revisited,''
  Phys.\ Rev.\ D {\bf 54} (1996) 2182.
  %%CITATION = HEP-PH 9602323;%%
 

 \bibitem{WW}  
  S.~Wandzura and F.~Wilczek,
  Phys.\ Lett.\ B {\bf 72}, 195 (1977).
  %%CITATION = PHLTA,B72,195;%%
 


\bibitem{An01}
I.~V.~Anikin and O.~V.~Teryaev,
  %``Wandzura-Wilczek approximation from generalized rotational invariance,''
  Phys.\ Lett.\ B {\bf 509}, 95 (2001)
  [arXiv:hep-ph/0102209].
  %%CITATION = HEP-PH 0102209;%%
  

  \bibitem{BL}
   G.~P.~Lepage and S.~J.~Brodsky,
  %``Exclusive Processes In Perturbative Quantum Chromodynamics,''
  Phys.\ Rev.\ D {\bf 22} (1980) 2157.
  %%CITATION = PHRVA,D22,2157;%%



\bibitem{DGP}
M.~Diehl, T.~Gousset and B.~Pire,
Phys.\ Rev.\ D {\bf 62}, 073014 (2000).
%%CITATION = HEP-PH 0003233;%%  
   

\end{thebibliography}
\end{document}